# Physics and the Totalitarian Principle

Helge Kragh[*]

**Abstract:** What is sometimes called the "totalitarian principle," a metaphysical doctrine often associated with the famous physicist Murray Gell-Mann, states that everything allowed by the laws of nature must actually exist. The principle is closely related to the much older "principle of plenitude." Although versions of the totalitarian principle are well known to physicists and often appear in the physics literature, it has attracted little reflection. Apart from a critical examination of the origin and history of the totalitarian principle, the paper discusses this and the roughly similar plenitude principle from a conceptual perspective. In addition it offers historical analyses of a few case studies from modern physics in which reasoning based on the totalitarian principle can be identified. The cases include the prediction of the magnetic monopole, the hypothesis of radioactive protons, and the discovery of the muon neutrino. Moreover, attention is called to the new study of metamaterials.

## 1. Introduction

Among the class of metaphysical principles which play a heuristic role in science, the simplicity principle also known as Occam's razor is a commonplace among philosophers and scientists alike. The subject of this paper is a much less-known principle or doctrine often referred to as the *totalitarian principle*, in the following abbreviated TP. According to this principle, what is allowed in nature is compulsory, that is, must necessarily happen or be the case. Another formulation is that what is possible must really exist. Although the TP only turned up in physics in the early 1960s, principally in areas of quantum and particle physics, in its spirit it can be traced back to Plato and other philosophers in ancient Greece. In the history of ideas the belief that all possibilities (in the form of potential existence) are actualized (in the form of real existence) is known as the *principle of plenitude* or PP for short.

The primary aim of this paper is to elucidate the TP and to call attention to how it sometimes enters physicists' way of thinking in the form of plenitude-related arguments. There is a great deal of confusion in how physicists refer to the TP and for this reason I start in Section 2 with a historical analysis of the origin of the TP in which I point out various mistakes appearing in the literature. Section 3 outlines the historically important but conceptually problematic PP and how it relates to the TP. One of the modern research areas I briefly discuss in this context is the interdisciplinary study of metamaterials, a study apparently far away from philosophy and yet one of considerable philosophical relevance.

[*] Niels Bohr Institute, Copenhagen University. E-mail: helge.kragh@nbi.ku.dk.

Apart from discussing the two meta-scientific principles in general terms, the paper also demonstrates by means of examples how this kind of thinking has played a role in the development of modern physics. Section 4 reconsiders Dirac's famous predictions of the antielectron and the magnetic monopole in the light of the PP, and in Section 5 I briefly discuss quantum mechanics in the same context. The sections 6 and 7 are devoted to cases of high-energy physics, the one dealing with muon physics and the other with theories assuming the proton to be radioactive.

The chosen cases are to some extent arbitrary, as they could have been supplemented or replaced with other TP-relevant cases. Some of these other cases come from chemistry (see Kragh 2019) and some come from astrophysics and cosmology. The recent idea of the multiverse is a case in point. According to George Gale (1990, p. 195), if "the well-known quantum theoretical dictum that 'whatever can happen, will'" is applied to the infinite multiverse, then all the niches in "cosmological ecology" are ensured to be filled with some world or other. The ontology of modern multiverse physics is in part inspired by reasoning associated with the PP and the TP. More can be said about this and other aspects of cosmology, but the subject will not be covered in this paper.

## 2. The totalitarian principle

In a couple of recent papers on the concept of randomness in quantum mechanics Gregg Jaeger (2016; 2017) has discussed not only the PP but also the stronger proposition that "everything not forbidden by law is compulsory." He incorrectly states that the proposition as well as the term "totalitarian principle" for it was originally formulated by Murray Gell-Mann in a paper of 1956.

With or without referring to the historical origin of the term, dozens of physicists have since the 1960s, and especially in popular and semi-popular writings, referred to the principle. As Werner Israel (1996, p. 607) observed, "'What is not forbidden is compulsory' is a saying well known among physicists." There is in the physics literature much confusion and carelessness with respect to the origin of the TP. It is often used without any reference, sometimes ascribed to Gell-Mann and at other times to the British novelist T. H. White. Before proceeding to the connection between the TP and the PP, and to a discussion of the status and use of these principles, it will be useful to present a brief history of the term TP.

First, with regard to Gell-Mann's alleged paternity to the PT, it is essentially a myth. In 1956 Gell-Mann wrote a paper which included an interesting footnote in which he introduced what he called the "principle of compulsory strong interactions," but this was not what came to be known as the TP. Here is what Gell-Mann (1956, p. 859) wrote:

> Among baryons, antibaryons, and mesons, any process which is not forbidden by a conservation law actually does take place with appreciable probability. We have made liberal and tacit use of this assumption, which is related to the state of affairs that is said to prevail in a perfect totalitarian state. Anything that is not compulsory is forbidden. Use of this principle is somewhat dangerous, since it may be that while the laws proposed in this communication are correct, there are others, yet to be discussed, which forbid some of the processes that we suppose to be allowed.

The first sentence comes close to the TP, but it is restricted to strong interactions and not stated as a general principle. Not only does the term TP not appear in the 1956 paper in *Nuovo Cimento*, it also does not appear in any of Gell-Mann's publications whether scientific or popular. He was undoubtedly aware of his alleged paternity, but chose never to comment on it.

The statement Gell-Mann associated with a totalitarian state is not what is usually known as the TP. On the contrary, it is the converse of it. This statement – that what is not compulsory is forbidden – can be found earlier in contexts which have nothing to do with either physics or philosophy. As pointed out by Edward Harrison, the phrase appears in an article in *Saturday Evening Post* of 23 December 1939 about Mussolini's Italy written by the journalist John T. Whittaker. In this article we read, "Coffee is forbidden, the use of motorcars banned and meat proscribed twice a week, until one says of Fascism, 'Everything which is not compulsory is forbidden'." Harrison (2000, p. 268) observes: "This is the principle of prohibition that attends authoritarian rule. The inverse is the principle of plenitude: everything that is not forbidden is compulsory. Whatever is possible must exist. In science we see that nature is plenitudinous rather than parsimonious."

If the TP cannot be found in Gell-Mann's works, could he have borrowed it from a literary source, in the same way that he borrowed the word "quark" from James Joyce's *Finnegan's Wake*? In a biography of Gell-Mann, George Johnson (1999, p. 224) claims that he was inspired by and borrowed the TP phrase from George Orwell's famous dystopian novel *Ninety Eighty-Four*. Likewise, admitting that "although the totalitarian principle is indisputably attributed to Gell-Mann, I could not trace the original source," G. F. Giudice (2008, p. 166) says that the TP "is actually coming from 'Nineteen Eighty-Four' by G. Orwell." Unfortunately there is no such phrase or anything close to it in Orwell's novel.

On the other hand, as pointed out by George Trigg (1970) in a letter in *Physics Today*, the phrase "anything not forbidden is compulsory" can be found in the British author Terence H. White's Arthurian novel *The Sword and the Stone*. This delightful story of Arthur's childhood was originally published by Collins in England in 1938 and, in a substantially revised version, by Putnam & Sons in the United States in 1939, but none of these editions contains the TP phrase. This was only the case in yet another revised version, which was eventually incorporated as book 1 in White's successful *The Once and Future King* from 1958. In the earlier versions the legendary sorcerer Merlin, acting as a tutor to young Arthur, had transformed the future king into a grass snake; but in the final 1958 version he turns instead the boy into an ant in order that he can enter an ant-hill (Brewer 1993).

An ardent pacifist, White pictured the ant colony as a totalitarian community populated by robot-like ants. The relevant passage is this (White 1958, p. 121):

> The fortress was entered by tunnels in the rock, and, over the entrance to each tunnel, there was a notice which said: "EVERYTHING NOT FORBIDDEN IS COMPULSORY." He read the notice with dislike, though he did not understand its meaning. … For some reason the notice gave him a reluctance to go, making the rough tunnel look sinister. He waved his antennae carefully, considering the notice, assuring himself of his new senses, planting his feet squarely in the insect world as if to brace himself.

There is no indication that White paid special attention to the later so oft-repeated phrase or that he thought of it in either a physics or philosophy context. To him it was just a notice related to the ant-hill story and it only appeared in this specific context. Yet, in many later references to the TP it is taken for granted that Gell-Mann borrowed the phrase from White; or it is simply ascribed to White, in some cases with full but wrong references to the early editions of *The Sword and the Stone*. To mention a recent example, Michael Turner (2015) cites Gell-Mann's dictum, which he had "borrowed from novelist T. H. White." Since Gell-Mann's paper appeared in 1956 and thus two years before White published the totalitarian phrase, obviously the physicist did not borrow it from the novelist.

The association between Gell-Mann and the TP may have arisen by a too generous reading of Gell-Mann's paper of 1956. Or perhaps the association is due to Gell-Mann's later prediction of the $\Omega^-$ particle and his quark model of strongly interacting particles (Bangu 2008). Quarks are theoretically allowed, hence they must exist. However, had Gell-Mann been inspired by some version of the TP he would expectedly have interpreted quarks as really existing particles emerging from what were formally un-actualized possibilities. In fact he did not. Gell-Mann initially emphasized that the quark model was a formal scheme and that real quarks did not exist. For several years he thought that quarks were "fictitious" and not real detectable particles (Johnson 1999).

The first time that the TP appears as a general "totalitarian" rule of physics seems to be in an essay of 1960 by Mael Melvin, a physicist from Florida State University, who referred to Gell-Mann's paper and offered his own version of the TP (Melvin 1960, p. 481). Three years later the TP was publicized in what became one of its standard forms, but in this case without citing Gell-Mann and without use the label "totalitarian." In a widely read popular book on particle physics Kenneth Ford (1962a, pp. 82-84) stated that "any behavior not prohibited by conservation laws will, sooner or later … actually occur." He described this "new view of democracy in nature" as a drastic change in the concept of natural law, since the essence of a fundamental law now became prohibition and not permission: "A conservation law is, in effect, a law of prohibition."

In a follow-up article on magnetic monopoles, Ford (1963b) referred to situations where experimenters had utterly failed to find any sign of a hypothetical particle and where the theorists on the other hand, "have failed to find any good reason why it should not exist." Such a situation he considered an "excellent reason" for maintaining interest in the particle, because

> One of the elementary roots of nature is that, in the absence of law prohibiting an event or phenomenon it is bound to occur with some degree of probability. To put it simply and crudely: Anything that *can* happen *does* happen.

Ford presumably came to this formulation of the TP independently. As he pointed out, the statement does not mean that scientists should be concerned about any conceivable non-existing features in nature, say hunting for laws to account for the fact that the Earth has only one moon and Venus has none. The number of moons around the planets is a contingent feature that may or may not be of scientific interest, but it is not governed by any law of nature.

By 1970 the TP had taken root in parts of the physics community, in particular in connection with unconfirmed but theoretically possible objects and processes. The principle was highlighted in papers on tachyons, magnetic monopoles, pulsars and gravitational collapse. For example, at the time it had been clearly demonstrated that the hypothetical tachyons are consistent with the theory of relativity, and for this reason they attracted much interest. Tachyons, if they existed, would possess strange and even seemingly paradoxical properties, but this was not enough to rule them out as possible real particles. In a paper of 1969 two physicists explicitly related the search for tachyons, and also for magnetic monopoles, to what they called Gell-Mann's totalitarian principle (Bilaniuk and Sudarshan 1969):

> There is an unwritten precept in modern physics, often facetiously referred to as Gell-Mann's totalitarian principle, which states that in physics "anything which is not prohibited is compulsory." Guided by this sort of argument we have made a number of remarkable discoveries, from neutrinos to radio galaxies. … Because theory does not exclude the possibility that a magnetic analog to the electric charge can exist, physicists persist in their quest for the magnetic monopole. … If tachyons exist, they ought to be found. If they do not exist, we ought to be able to say why not.

Apart from particle physics, astrophysics was another research area used to illustrate the relevance of the TP. Reflecting on the recent discovery of pulsars and their explanation in terms of spinning neutron stars, Jeremiah Ostriker (1971) praised the rule that "anything not forbidden is compulsory," which he associated with White rather than Gell-Mann. As he saw it, the rule or principle had predictive power as it had led to the discovery of improbable objects as diverse as neutrinos and pulsars.

## 3. The principle of plenitude

Although the TP is of recent vintage, its essential message of equality between potential and actual existence can be found much earlier. The TP can reasonably be regarded as a physics version of the age-old metaphysical idea which Arthur Lovejoy coined the "principle of plenitude" (Lovejoy 1936; Kragh 2019). However, Lovejoy carried on his erudite study only to the romantic philosophy of nature in the early nineteenth century. Moreover, he focused on the life sciences and had almost nothing to say about the physical and mathematical sciences. Readers of *The Great Chain of Being* will look in vain for the names of Newton, Euler, Priestley, Laplace and Lavoisier.

The later literature on plenitude reasoning is predominantly philosophical or written in the same tradition of history of ideas, and with the same limitations, as Lovejoy's work. Only a few scattered studies pay attention to how the PP or something like it has influenced ideas and theories in the more recent era of science and even fewer are concerned with the physical sciences. From a philosophical perspective Robert Kane (1976) was the first to consider the PP in relation to modern physics and suggest cases where the principle has played a heuristic role. As regards the physicists, they seem generally to be unaware of the PP. This is evidenced by a search for "principle of plenitude" in the widely read publications *American Journal of Physics*, *Reviews of Modern Physics* and *Physics Today*. The search yields no results.

As a meta-scientific postulate the PP shares its status with other and better known postulates such as the principles of simplicity and unity. But the spirit of plenitude philosophy is hard to reconcile with the other two principles. As noted by Alan Baker (2007, p. 199), "According to Occam's Razor we ought *not* to postulate the existence of unicorns. According to the principle of plenitude we *ought* to postulate their existence." Again, whereas the PP highlights nature's amazing richness and diversity, in its ontological version the principle of unity goes in the opposite direction. Physicists have often and mostly successfully been guided by their desire to reduce the diversity of matter to as few buildings blocks as possible, ultimately to a single one. Far from seeking ontological plenitude they seek ontological paucity.

The PP in its classical version states that what is conceived as possible must also have a physical existence and thus belong to the real world. Or phrased differently, what can exist does exist. As regards the premise, the criterion for what can exist or is possible includes two elements. The object or phenomenon in question must be allowed, firstly, by reasons of logic and, secondly, by the fundamental laws of nature. Something which is internally inconsistent cannot exist either potentially or actually (and yet it can sometimes be imagined). Whereas consistency is thus a necessary condition, it is not a sufficient condition. It is of course possible that something exists even though it is ruled out according to the known laws of nature.

If evidence is strong enough the anomalous object or phenomenon is admitted as real, with the consequence that one or more of the laws of nature are in need of revision.

As to the term "possible" it can be understood in different ways, but in so far that physicists are concerned, they tend to judge something to be possible if and only if it is compatible with the laws of nature whether or not these laws are currently known. In normal scientific practice the existence or non-existence of some hypothetical object or phenomenon will be judged according to the empirical evidence for and against the object. However, in its pure form the principle of plenitude is not concerned with evidence, as it is enough that the object is possible, that is, conceivable under the constraints of the laws of nature. The object in question may have extremely bizarre and wildly implausible properties, but again this is irrelevant in a plenitude context. On the other hand, it is relevant if the properties turn out to be logically contradictory.

Plenitude reasoning involves a historical element in so far that what is considered physically or genuinely possible depends on the best scientific knowledge at any given time. What is scientifically allowed at one time may be forbidden at a later time, or vice versa. For example, according to the electron theory of the early twentieth century the charge of the positive elementary particle (the proton) could have any value relative to the electron's numerical charge. Consequently, an ordinary hydrogen atom with a charge excess was a theoretical possibility. Nor were there any reasons to exclude sub-electrons from nature's fabric or, for that matter, electrons with much greater mass than the one observed (Holton 1978). From a plenitude point of view one should expect such particles to exist. However, with the development of the extended electroweak theory in the 1970s sub-electrons and hydrogen atoms with a charge excess were theoretically forbidden and thus no longer candidates for potential existence.

As Lovejoy pointed out, there is more than one version of the PP. The original one advocated by Spinoza and Leibniz was static, presupposing that the actualization of possibilities was independent of time. But with the growth of evolutionary ideas in natural philosophy in the late eighteenth century, the PP came to be understood as a temporal principle postulating that at *some time*, and not necessarily at the present, all possibilities must be realized (see also Kane 1976). In other words, the temporal version of the what-can-exist-does-exist formula may refer to any time in the future or to any time in the past. This alone makes the formula untestable. Moreover, the temporally extended meaning is supplemented with a spatial extension, in the sense that the realization of a possibility can occur anywhere in the universe, at *some location* which does not need to be the Earth.

Existence refers to nature, which traditionally meant the surface of the Earth. On the other hand, ever since the mid-nineteenth century scientists have produced in their laboratories a large number of objects with no known counterpart in undisturbed nature. Would we say that something synthesized in the laboratory and detected only there (such as superheavy chemical elements), exists in the same sense that an entity found outside the laboratory? According to the traditional meaning of the PP the answer will be no, but according to the modern TP it will be yes. Anti-hydrogen atoms and anti-helium nuclei have not been found in nature, but the short-lived objects have been detected in the laboratory and so they unquestionably exist. Generally speaking it is problematic to maintain a strict division between nature and the laboratory when it comes to the existence of objects or processes. The reason why anti-hydrogen atoms do not exist naturally is not that they are forbidden by law but rather that the natural environment is unfriendly to such atoms.

The modern study of metamaterials has not as yet attracted philosophical interest but it is of relevance to the PP and also to the philosophy of science and technology more generally. Metamaterials are complex materials engineered or theoretically conceived in such a way that they have unusual electromagnetic and other properties not found in naturally occurring materials, such as a light-transmitting substance with negative refraction index (Beech 2012, pp. 131-187). The possibility was theorized by the Russian physicist Victor Veselago in the late 1960s, and some forty years later the first material with a negative refraction index was constructed. The potential existence of metamaterials seems to correspond to their actual existence.

In a discussion of the meaning of metamaterials Ari Sihvola (2002, p. 16) has argued that although in many cases they have properties not found in nature they nonetheless are natural. His argument is that they are designed and synthesized in accordance with the known laws of nature. Moreover, without referring to either the PP or the TP he expresses himself in complete agreement with these principles:

> If we did not find materials with properties corresponding to these designs yet in nature, is it only that we did not search hard enough? Nature is extremely prodigious. All possibilities come through. All properties, all media that are not forbidden, are there, somewhere. They are compulsory.

The philosophy of optical metamaterials does not focus on why things are as they are, but rather on why unknown things do not exist or occur (Cai and Shalaev 2010, p. vii). There are in the scientific literature on metamaterials other implicit references to the PP and in one case plenitude rhetoric appears explicitly (Beech 2012, p. 131):

> "Everything not forbidden is compulsory." So writes T. E. White in his book *The Once and Future King* – and, if nature ever had need of a motto then surely it would be these words. For indeed, nature is the ultimate utilitarian, ever building, ever moving, never stationary, searching in all possible ways, step by small step, for a better adapted, more efficient, more successful survivor. If a process is not physically impossible, that is, against the laws of physics, nature will usually find a use for it. Metamaterials, however, appear to be one trick that nature has missed – so far at least.

In fact, there are many more tricks that nature has missed and which disagree with the TP applied to undisturbed nature.

## 4. Magnetic monopoles reconsidered

The best known episode in the history of modern physics illustrating plenitude reasoning or the use of the TP is probably Paul Dirac's prediction in 1931 of the magnetic monopole (Kragh 1990, pp. 272-274). In a nutshell, in his remarkable paper of that year Dirac concluded that magnetic poles corresponding to electrical charges could be described within the framework of quantum electrodynamics. Although Dirac (1931) knew that such particles were purely hypothetical, he stressed that quantum mechanics "leads inevitably to wave equations whose only physical interpretation is the motion of an electron in the field of a single [magnetic] pole." Adopting a weak form of the plenitude argument he suggested that since there was no theoretical reason barring the existence of monopoles, they probably existed somewhere in nature. "Under these circumstances," he wrote, "one would be surprised if Nature had made no use of it." Notice that Dirac did not appeal to the more rigid TP formula where allowed entities are compulsory. His more cautious version and the one generally adopted by physicists can be stated as "that which is not forbidden is allowed and hence expectedly exist."

In his paper of 1931 Dirac also predicted the antielectron as a new particle soon to be known as the positron. The previous year, at a time when he still believed that the antielectron was a proton, he proposed proton-electron annihilation ($p^+ + e^- \rightarrow 2\gamma$) with the sole argument, "There appears to be no reason why such processes should not actually occur somewhere in the world. They would be consistent with all the general laws of Nature" (Dirac 1930). In the case of the proton as an antielectron, he was not guided so much by plenitude reasoning as by his belief in simplicity and a unity of nature. As William Prout in 1815 had suggested all matter to consist of hydrogen atoms, so Dirac suggested that the electron was the only building block of matter. In 1931, after having abandoned what he called "the dream of philosophers," he thought that it might be possible to create antielectrons by photon-photon collisions in the laboratory ($\gamma + \gamma \rightarrow e^- + e^+$). On the other hand, "we should not expect to find any of them in nature."

In 1932 the positive electron turned up in nature, but for a while there was a great deal of confusion whether or not the observed particle was the same as the one predicted by Dirac. The two predictions of antiparticles dating from 1930 and 1931 had in common that they were based on Dirac's picture of the vacuum as consisting of unoccupied positive-energy states plus an infinite uniform distribution of negative-energy electrons governed by Pauli's exclusion principle. Only the vacancies or holes in this world of negative energy would be observable and hence count as real particles. In a way, Dirac assigned real existence to what was formally potential existence. This has made two philosophers of physics to suggest that the prediction of antielectrons was strongly guided by plenitude reasoning, not in the sense of the TP but in the original PP sense related to the idea of a great chain of being (Massimi and Redhead 2003):

> They were the necessary – until then, missing – link of a sort of great chain of being (echoing Lovejoy's famous expression) going from a lower to an upper bound of a uniformly distributed and filled continuum of energy states: they were introduced by the *principle of plenitude* and *continuity* to fill up the few unoccupied links of an infinite chain.

To return to the monopoles, given that they were allowed and yet absent in nature Dirac had to find a reason for their non-existence, which he found in the strong coupling between oppositely charged magnetic poles. Dirac's monopole theory only attracted wide interest in the 1960s, when physicists began looking for the hypothetical particles in experiments and developing the theory in various directions (Ford 1963b). Since then thousands of papers have been written on the still undetected magnetic poles. In a review paper of 1970 a team of American physicists concluded from experimental data, that *if* monopoles exist they must be extremely rare. They considered Dirac's plenitude argument to be "particularly relevant if Gell-Mann's statement is valid that what is not expressly forbidden is 'obligatory'" (Fleischer et al. 1970). Statements like this still occasionally appear in the research literature. Kimball Milton (2006, p. 1639) characterized Dirac's prediction as "an example of what Gell-Mann would later call the 'totalitarian principle' – that anything which is not forbidden is compulsory."

Whether in Dirac's original version or in the later version based on grand unified theories, where magnetic monopoles are inevitable, to this day the elusive particles remain unobserved. There may not be a single monopole in the visible universe and yet, as is widely believed, monopoles may have been created abundantly during the early phase of the universe. If so they exist in the temporal sense of the PP, illustrating that "It is only of the universe in its entire temporal span that the principle of plenitude holds good" (Lovejoy 1936, p. 244). Similarly, although free quarks have never been detected, according to the standard cosmological model they existed from about $10^{-12}$ s to $10^{-6}$ s after the big bang when nucleons had not yet been formed. From the perspective of the temporal PP, the free quarks of the cosmic past count as realizations of possible particles.

## 5. Quantum processes

As a general ontological postulate the TP is not restricted to modern physics. Nonetheless, it is often associated with the probabilistic indeterminism characteristic of quantum-mechanical processes, where it appears in a new light. When Ford (1963b) stated that allowed events must occur "with some degree of probability," he was thinking of quantum mechanics.

More specifically Richard Feynman's path integral formulation of quantum mechanics dating from 1948, also known as the sum over histories formulation, can be taken as an illustration of the TP or the PP (Krieger 1992, p. 62). According to Feynman, a process going from an initial space-time state A to a final state B can proceed by a number of different paths or "histories." We have to consider *all* possible paths and associate each of them with a certain amplitude and probability. When all the probabilities are superposed, the sum gives the probability of the transition from A to B. Although the resulting path is the only one observed, the other possible paths are still necessary parts of the superposition. They are potential and yet make themselves felt in the outcome. Feynman later commented on his path integral approach to quantum mechanics as follows (Feynman, Leighton and Sands 1966, p. 19-9):

> Is it true that the particle doesn't just "take the right path" but that it looks at all the other possible trajectories? … The miracle of it all is, of course, that it does just that. That's what the laws of quantum mechanics say. [The principle of least action] isn't that a particle takes the path of least action, but that it smells all the paths in the neighborhood and chooses the one that has the least action.

As to the possible paths or histories it is generally held that they just enter computationally, without implying that they correspond to actual paths. But from a strict plenitude point of view it would be natural to ascribe them reality, such as some philosophers have recently suggested (Terekhovich 2018).

Probability is not necessarily a measure of possibility, as the two terms can be interpreted in a variety of ways. Still, most physicists undoubtedly share the opinion of Mario Bunge that probability warrants possibility whereas the opposite is not the case. Taking quantum and other probabilistic processes into consideration, Bunge (1976, p. 31) suggested a more refined version of the totalitarian principle, namely "all consistent repetitive chance events (in particular all those consistent with the conservation laws) are likely to occur in the long run."

## 6. Weak interactions

In late 1930 Wolfgang Pauli famously suggested that the problem of beta decay might be solved if the radioactive nucleus not only emitted an electron but also an accompanying neutrino, or what he at the time called a "neutron." Contrary to what has been claimed (see Section 2), Pauli's proposal was not motivated by TP-like reasoning. He originally justified the proposal in terms of conservation laws without claiming that the neutrino really existed. In a letter to Oskar Klein of 12 December 1930 he wrote, "If the neutrons [neutrinos] really existed, it would scarcely be understandable that they have not yet been observed; for this reason, I also do not myself believe very much in the neutrons" (Pauli 1985, p. 45).

Although Pauli realized that there was no direct experimental evidence in favour of the neutrino, within a year or so he came to think that it was after all a real particle. But he did not think so just because it was allowed by the laws of fundamental physics. To him, this was a necessary but not sufficient reason to believe in the reality of the new particle.

The neutrino associated with beta decay was at first controversial. But by the late 1930s it had become generally accepted, many years before it was detected by Frederick Reines and Clyde Cowan in experiments of 1956. Within a few years it turned out that the Pauli-Fermi electron neutrino was not the only of its kind. In 1988 the Nobel Prize was awarded to Leon Lederman, Melvin Schwartz and Jack Steinberger at the Brookhaven National Laboratory for having proved, in accelerator experiments of 1962, the existence of the muon neutrino $\nu_\mu$. The three physicists studied the decay $\pi \to \mu + \nu_\mu$ and established that there are at least two lepton families. As $(e, \nu_e)$ formed a pair, so did $(\mu, \nu_\mu)$.

The discovery of the muon neutrino provides a possible case of PT-related research. An important reason for the hypothesis of a new kind of neutrino was unobserved processes which ought to exist since they satisfied all known conservation laws, an example being the decay $\mu^\pm \to e^\pm + \gamma$. All theories predicted that this process should accompany the normal muon decay $\mu^\pm \to e^\pm + \nu + \bar{\nu}$ to a ratio of $1:10^4$, but precise experiments failed to detect just a single process with the $\nu + \bar{\nu}$ annihilation signature. Why?

As Lederman (1963a) put it in a lecture shortly after the discovery, "It conserves everything you can think of, and yet it is not observed. Gell-Mann's theorem, which is the totalitarian theorem, says that in physics anything that is not forbidden is compulsory. This is one reason why people were disturbed at not seeing this reaction." Likewise in a paper in *Scientific American*, where Lederman (1963b) did not quote the TP but paraphrased it: "When reactions that could take place are not seen, one must conclude that a basic prohibition law is at work." In the case referred to the reason for the non-existence of neutrino annihilation was that the two neutrinos are not in fact particle and antiparticle but belonging to different species, one being an electron neutrino and the other a muon neutrino. It is not enough that the lepton number is conserved, for the muon number and the electron number must also be conserved separately. This is the prohibition law alluded to by Lederman.

Notice that here the TP operates in the negative sense. Not: since X is possible, X must exist; but: X does not exist, so there must be a reason or law that warrants the non-existence of X. Generally, TP-reasoning invites physicists to focus on processes that *do not* occur rather than those that *do* occur. In this way they have been able to establish a number of partial and absolute conservation laws.

## 7. Proton decay

Another example illustrating the TP in particle physics is provided by the lack of spontaneous proton decay. As early as 1938 Ernst Stueckelberg suggested that the non-occurrence of protons decaying into lighter particles could not be understood on the basis of known conservation laws: "No transmutations of heavy particles (neutron and proton) into light particles have yet been observed in any transformation of matter. We shall therefore demand a conservation law of heavy charge" (Pais 1986, p. 488). Persistent failures in detecting proton decay, and also free neutron decay into leptons, led in the 1950s to the law of baryon conservation forbidding processes such as $p^+ \rightarrow e^+ + \pi^0 \rightarrow e^+ + 2\gamma$. The discovery of baryon conservation may be seen as one more example of the heuristic use of the TP in its negative version: where a process or an object is never realized whether in experiments or in nature, there must be a compelling reason for it, namely a law forbidding the process or object. Similar reasoning was behind the idea of a conserved lepton number dating from 1953 and ruling out, for example, the process $\bar{\nu}_e + n \rightarrow p^+ + e^-$.

Baryon conservation is part of the standard model but not of extensions of this model (grand unification, supersymmetry), which on the contrary require the proton to be an unstable particle. Grand-unified calculations of the late 1970s predicted the proton to be unstable with an extremely long half-life, which today is known to be longer than ca. $10^{34}$ years. In an early review of the possibility of unstable protons, Steven Weinberg (1981, p. 64) pointed out that, in a sense, protons ought to decay, because "Experience in the physics of elementary particles teaches that any decay process one can imagine will occur spontaneously unless it is forbidden by one of the conservation laws of physics." This is an argument close to the TP except for the problematic and probably carelessly used word "imagine." After all, the TP does not state that what is imaginably possible exists, but only that what is genuinely or physically possible exists.

Different theories of fundamental physics lead to different conceptions of what is forbidden or not, and therefore also to different conceptions of what is "compulsory" or not. Leonard Susskind (2006, p. 189) assumed a version of grand unification to be true when he, in a popular book on string theory and multiverse physics, referred to proton decay in terms of the TP:

> As Murray Gell-Mann once quoted T. H. White, "Everything which is not forbidden is compulsory." Gell-Mann was expressing a fact about quantum mechanics … [that it] will eventually make anything happen unless some special law of nature forbids it. … It ought to be possible for protons to disintegrate into photons and positrons. No principle of physics prevents it.

As pointed out, the first sentence in the quotation is erroneous.

## 8. Conclusions

In this essay I have discussed the meaning and use of the TP, in part in a general perspective and in part as related to specific cases in the history of physics. I have suggested that the TP is the successor of the old and venerable PP specially adapted to modern physics. The TP emphasizes agreement with the laws of physics as the sole criterion of actual existence. This is a postulate that on occasion works as a heuristic tool, as one motivation among others for scientists to explore if something theoretically allowed is in fact part of nature's fabric. Thus, plenitude reasoning may generate hypotheses, but it is generally accepted that the truth of these hypotheses can only be ascertained by empirical investigation. A plenitude-generated hypothesis unsupported by empirical evidence will not survive for long or be taken very seriously. The relevant empirical evidence is sometimes evidence that something exists, whereas in other cases it is evidence that something which ought to exist does not exists. In modern physics the latter kind of evidence is no less important than the former kind.

As a statement of what exists in nature the TP lacks conviction. There are numerous non-forbidden objects and processes that simply do not exist, at least not in our universe. Thus, complex atoms made up of antiparticles satisfy the TP and yet they do not exist in nature. There may be reasons for their non-existence, but anti-atoms are allowed by the known laws of physics. Physicists have created minute amounts of anti-hydrogen in the laboratory and in this sense these simple anti-atoms do exist. Generally, the laboratory synthesis of non-natural objects invites a formulation of the TP broader than the one found in the classical PP. It is no longer a claim limited to what exists in nature at some time, but a claim of what exists in nature or can be created in the laboratory.

Although references to the TP in one form or other are fairly frequent in the physics literature, they are mostly implicit and restricted to popular and semi-popular journals or, in a few cases, to conference proceedings. In the few cases where the TP appears in research papers, it is always as an afterthought and without any significance for the content of the paper. As pointed out by Sorin Bangu (2008, p. 251) in an analysis of Gell-Mann's prediction of the omega-minus particle, the TP and similar precepts "have always had an anecdotal value in the physics community and, significantly, they only appear mentioned in the popular ('philosophical') presentations of scientific results." The TP has become part of physicists' folk history but is rarely taken seriously as a scientific principle. In this respect it differs from other principles discussed by physicists, such as, for example, the cosmological principle, the equivalence principle or even the controversial anthropic principle.